\input harvmac


%
%
\ifx\epsfbox\UnDeFiNeD\message{(NO epsf.tex, FIGURES WILL BE IGNORED)}
\def\figin#1{\vskip2in}
\else\message{(FIGURES WILL BE INCLUDED)}\def\figin#1{#1}\fi
\def\ifig#1#2#3{\xdef#1{fig.~\the\figno}
\goodbreak\midinsert\figin{\centerline{#3}}%
\smallskip\centerline{\vbox{\baselineskip12pt
\advance\hsize by -1truein\noindent\footnotefont{\bf Fig.~\the\figno:}
#2}}
\bigskip\endinsert\global\advance\figno by1}

\def\ifigure#1#2#3#4{
\midinsert
\vbox to #4truein{\ifx\figflag\figI
\vfil\centerline{\epsfysize=#4truein\epsfbox{#3}}\fi}
\narrower\narrower\noindent{\footnotefont
{\bf #1:}  #2\par}
\endinsert
}


\def\IC{{\ \hbox{{\rm I}\kern-.6em\hbox{\bf C}}}}
\def\IR{{\hbox{{\rm I}\kern-.2em\hbox{\rm R}}}}
\def\IZ{{\hbox{{\rm Z}\kern-.4em\hbox{\rm Z}}}}

\def\sIR{{\hbox{{\sevenrm I}\kern-.2em\hbox{\sevenrm R}}}}

\Title{RU-97-105, UTTG-27-97, PUPT-1755}
{\vbox{\centerline{Evaporation of Schwarzschild 
Black Holes in Matrix Theory }}}

\centerline{\it T. Banks~$^1$, W. Fischler~$^2$,  I.R. Klebanov~$^3$ 
}
\medskip
\centerline{$^1$Department of Physics and Astronomy}
\centerline{Rutgers University, Piscataway, NJ 08855-0849}
\centerline{\tt banks@physics.rutgers.edu}
\medskip
\centerline{$^2$Theory Group, Department of Physics}
\centerline{University of Texas, Austin, TX 78712}
\centerline{\tt fischler@physics.utexas.edu}
\medskip
\centerline{$^3$Joseph Henry Laboratories}
\centerline{Princeton University, Princeton, NJ 08544}
\centerline{\tt klebanov@puhep1.princeton.edu}

\bigskip

\medskip

\noindent
Recently, in collaboration with Susskind, we proposed
a model of Schwarzschild black holes in Matrix
theory. A large Schwarzschild black hole is described by a 
metastable bound state of a large number of
D0-branes which are held together by a background, whose structure
has so far been understood only in 8 and 11 dimensions.
The Hawking radiation proceeds by emission of small clusters 
of D0-branes. We estimate the Hawking rate in the Matrix theory model
of Schwarzschild black holes and
find agreement with the semiclassical rate up to an undetermined
numerical coefficient of order 1.

\hyphenation{Min-kow-ski}
\Date{December 1997}

\newsec{\bf Introduction}

In a recent paper \ref\mbh{T. Banks, W. Fischler, I.R. Klebanov,
L. Susskind, hep-th/9711005.} Susskind and the authors
presented a model of Schwarzschild black
holes in Matrix Theory \ref\bfss{T. Banks, W. Fischler, 
S. Shenker, L. Susskind, Phys. Rev. D55
(1997) 112, \break hep-th/9610043. }.  
The key feature of the model was the
notion of a {\it Boltzmann gas of D0-branes} (which we will review briefly
below).\foot{For related work, see \ref\vol{
I.Volovich, hep-th/9608137.}, \ref\hm{G. Horowitz and E. Martinec,
hep-th/9710217.}, \ref\li{M. Li, hep-th/9710226.}, \ref\minic{
D. Minic, hep-th/9712202.} }  
We showed that the Matrix model for M theory in
$11$ noncompact dimensions, and also for its toroidal compactification 
to $8$ noncompact dimensions, contained a set of states
which obeyed (up to a numerical coefficient which we could not
calculate) the Bekenstein-Hawking relation between entropy and
transverse area for a Schwarzschild black hole.  
The relation between energy and entropy for these states also obeyed the
black hole formula.\foot{In the supersymmetric Yang-Mills
formulation on the dual torus the corresponding results were obtained in
\ref\bfks{T. Banks, W. Fischler, I.R. Klebanov, L. Susskind,
hep-th/9709091.}, \ref\ks{I.R. Klebanov, L. Susskind,
hep-th/9709108.}, \ref\dmrp{S. Das, S. Mathur, S.K. Rama and
P. Ramadevi, hep-th/9711003.}.}  
We were also able to compute the long distance Newtonian
gravitational force between equal mass black holes, with an answer in
agreement with classical gravity.

The purpose of the present note is to calculate the rate of Hawking
radiation from our model black holes.  In \mbh\ we showed that the
individual bound D0-branes in the model, had the kinematic properties of
Hawking radiation in the boosted frame in which we examine our black
hole.  The D0-branes are \lq\lq tethered\rq\rq\ to a classical
background by harmonic forces. 
In this note we argue that the probability for the
classical variables which produce these forces on an individual 
D0-brane to fluctuate to zero is independent of the mass of the black
hole in the large mass limit.  We show that when this
estimate is combined with the proper phase space integral, it gives a
decay rate for the boosted black hole which is just the Lorentz
transform of the rest frame Hawking evaporation rate.

To briefly summarize our black hole model: we consider the Hilbert space
of the matrix theory which represents Discrete Light Cone Quantized 
M theory compactified on some manifold $Y$ 
\ref\lennynatisen{L. Susskind, hep-th/9704080; N. Seiberg, 
Phys. Rev. Lett. 79 (1997) 3577, hep-th/9710009;
A. Sen, hep-th/9709220.} in the sector with $N$ units of
longitudinal momentum ($DLCQ_N$).  The radius of the
light-like circle will be denoted by $R$. 
We will also choose the noncompact
transverse spacetime dimension $D$ to be greater than or equal to $6$.
The model contains a set of variables which includes matrices $X^i$
representing the transverse positions of $N$ D0-branes in a weakly
coupled Type IIA string theory called 
\lq\lq the analog model.\rq\rq\  We emphasize that these 
are not the Boltzmann D0-branes of which our
black hole is constructed.  The matrices also describe creation and
annihilation operators for strings stretching between the D0-branes.

We consider a semiclassical configuration of the variables of the model,
which includes a semiclassical background $X^i_{cl}$ for the transverse
position matrices.   This background configuration must satisfy a number
of constraints, which were described in \mbh\ . 
Boltzmann D0-brane positions are defined as perturbations of the
background
\eqn\bzb{X^i_{cl} \rightarrow X^i_{cl} + \sum_{n=1}^N r^i_n \delta_n}
where the $\delta_n$ are a set of independent commuting matrices chosen
to minimize the quantity $\vert Tr [X^i_{cl} , \delta_n ]^2 \vert$.
This is a term in the matrix model Hamiltonian which gives rise to a
harmonic potential binding these Boltzmann D0-branes to the classical
configuration.  In \mbh\ we argued that for $D \geq 6$ we could choose
our classical configuration so that this harmonic potential did not
interfere with the scaling argument we review in the next paragraph.

In matrix theory, the Bose or Fermi statistics of particles arises as a
residual gauge symmetry.  Since the entire gauge symmetry of the model
is broken by the classical background, this means that the variables
$r^i_n$ should be treated as the coordinates of distinguishable, or
Boltzmann, particles.  The approximately degenerate configurations of
these particles then have an entropy of order $N$.  In \mbh\ we argued
that the effective Hamiltonian of these degrees of freedom gave rise to
bound states whose transverse area is of order $G_D N$ and whose light
cone energy is of order $N R (G_D N)^{- 2/(D-2)}$.  When translated into
invariant mass, this gives precisely the mass/entropy/area relations of
a Schwarzschild black hole.  We were unable to calculate the numerical
coefficients in these relations because we did not have the full
effective Hamiltonian of the Boltzmann gas\foot{Recently, Liu and
Tseytlin \ref\lt{H. Liu and A. Tseytlin, hep-th/9712063.} 
have proposed to describe the interactions of
the Boltzmann D0-branes by 
a Hamiltonian containing terms of all orders in
the velocity \ref\bbpt{K. Becker, M. Becker, J. Polchinski and
A. Tseytlin, Phys. Rev. D56 (1997) 3174, hep-th/9706072.}.
Perhaps this can be
used to make some progress on the numerical coefficients.} and because
the mean field approximation which we used was very crude.

According to the above discussion, the light cone energy per particle of
the Boltzmann gas is $\sim R (G_D N)^{- 2/(D-2)}$, where
$G_D$ is the Newton constant in $D$ non-compact dimensions.
The transverse momentum 
per particle is $\sim (G_D N)^{- 1/(D-2)}$ and 
the longitudinal momentum is just
$1/R$. We showed that these were precisely
the kinematical properties of the Hawking particles 
when boosted into a frame where the black hole has
longitudinal momentum $P_-=N/R$. 
This suggests the following attractive and simple
picture of the Hawking evaporation process: the classical background
provides a harmonic potential which binds the Boltzmann D0-branes to the
black hole.  The background itself should be represented as a coherent
quantum state centered around a periodic solution of the classical
equations of motion of the matrix model.  According to the wave function
of this state, there is a certain amplitude for the part of the
classical configuration which interacts with a given D0-brane, to
fluctuate to zero.  From the point of view of a basis in which the
particular D0-brane of interest (say $\delta_1$) occupies the lowest
right hand corner of the matrix, the relevant part of the classical
solution is the part which occupies the last row and column.  There are
thus $o(N)$ possible degrees of freedom which might be excited.

However, in the explicit classical backgrounds which we constructed in
$D=11,8$ \mbh, only $o(1)$ of these possible background degrees of
freedom were actually utilized.  In $D=11$, this is a consequence of
approximate locality of the action on the world volume of the classical
membrane.  In $D=8$ the classical background for a typical black hole
could be viewed as a lattice of D0-branes connected by strings on a
3-torus, with the lattice spacing of order $N^{-1/3}$ \bfks.
Energetic considerations ensured that
only strings attaching a given D0-brane to $o(1)$ nearest neighbors on
this lattice were excited.  We will assume that similar pictures work
for all $D \geq 6$.  

As a consequence, it seems reasonable to assume that the amplitude to
\lq\lq liberate a D0-brane\rq\rq\ from the black hole is independent
of $N$ as $N \rightarrow \infty$.  It is the value of the wave function
of the system on a submanifold of codimension which is $o(1)$.

\newsec{Calculation of the Hawking Evaporation Rate}

Given the estimate of the D0-brane liberation amplitude in the
previous section, we can proceed to calculate the Hawking rate.
The quantum fluctuation of the background described in the previous
section gives rise to a single D0-brane wave function which is, by the
estimates of \mbh, a smooth function $A(y)$ of rapid decrease in the variable
\eqn\scalvar{y = {\vert p_\perp\vert \over (G_D N)^{- 1/(D-2)} }
\ ,}
where $p_\perp$ 
is the transverse momentum and $G_D$ is the $D$-dimensional
Newton constant.  
Thus, the amplitude to produce a D0-brane with momentum
much larger than the Hawking momentum is highly suppressed.
If we assume that the fluctuations which liberate any of the $N$ 
D0-branes are independent and incoherent, then the probability per unit
time to emit a Hawking particle is given by
\eqn\proba{ {dN\over d x^+} \sim N {1\over R} \sum_{n>0}
\int_0^\infty dp_+  \int d^{D-2} p_\perp 
\delta \left ({n\over R} - {p_\perp^2\over p_+ }\right )  
\vert A(n,y)\vert^2 }
where $y$ is the variable defined above. In this equation we have
generalized our considerations to include processes in which a cluster
of $n$ D0-branes, with $n$ 
finite and independent of $N$, is emitted. Such systems also have the
kinematic properties of Hawking radiation.   However, since the cluster
of D0-branes is connected to the classical background by $o(n)$
degrees of freedom, we should expect the matrix element to fall off
exponentially in $n$.  The overall factor of $N$ in equation \proba\
represents the incoherent sum over processes in which a particular bound
D0-brane is liberated.  Finally, we have used relativistic phase space
to integrate over the final states of the outgoing D0-brane.  
Although our calculation is done in a particular frame, chosen so that
the geometrical structure of the black hole just fits inside the
$DLCQ_N$ quantization volume, we expect that the matrix element which we
have estimated is in fact, for large $N$, approximately the invariant S
matrix element of a Lorentz covariant system.  Of course, the
crucial, as yet unproven, assumption of Matrix Theory is that the S
matrix computed by the theory is indeed Lorentz invariant.  We cannot
prove this claim at present, nor can we show that our approximate
evaluation of the matrix element is accurate.  However, assuming that
these claims are valid, the correct rate is obtained by integrating our
matrix element against relativistic phase space.

Our measure thus contains an extra factor of light cone energy, $p_+$,
compared to the nonrelativistic phase space of the D0-brane quantum
mechanics. This factor, which endows the measure with
the correct boost transformation properties,
has been absorbed into the longitudinal momentum delta-function in \proba\ .

Finally, we need to estimate the square of the matrix element
for $n$ D0-branes to be liberated, $|A(n,y)|^2$. As we have explained,
this quantity is appreciable only if $n$ and $y$ are of order 1.
$|A(n,y)|^2$ has dimensions of length$^{D-2}$. The only dimensionful
quantities at our disposal are $R$, $l_{11}$ and the radii of the
compactification torus, $L_i$. Since the measure 
in \proba\ transforms properly
under boosts, any dependence of $|A(n,y)|^2$ on $R$ would violate
Lorentz invariance. Furthermore, we will assume that
the dependence on $l_{11}$ and the radii is through the $D$-dimensional
Newton constant only.\foot{This assumption is plausible
because $G_D$ is the only quantity that appears
in the D-brane interaction Hamiltonian, but its better justification is
clearly necessary.} Thus, we are led to
$$ |A(n,y)|^2 \sim G_D\ ,
$$
for $n$ and $y$ of order 1.
Estimating the Hawking rate \proba\ with this assumption, we find
\eqn\rate{ {dN\over d x^+} \sim R (G_D N)^{- 2/(D-2)}\ . }

By way of comparison, we now
compute the Hawking radiation rate according to
the conventional semiclassical formulae. 
First, let us write down the Hawking rate  in the usual equal-time
quantization.  The answer can be written as
\eqn\rest{ {dN\over d x^0} \sim 
\int_0^\infty dp_+ \int_0^\infty dp_- \int d^{D-2} p_\perp 
\delta (p_+ p_- - p_\perp^2)  
e^{-p_0/T_H} {\cal A} p_0
\ ,
}
where ${\cal A}= 4 G_D S$ is the horizon area. 
We have included the thermal factor appropriate for the Boltzmann
statistics.
The Hawking temperature is related to the Schwarzschild radius $R_S$
by
\eqn\temp{T_H\sim {1\over R_S} \ ,}
and
\eqn\rs{R_S \sim (S G_D)^{1/(D-2)} \sim (N G_D)^{1/(D-2)}
\ .}
An explicit factor of $p_0$ is
needed in \rest\ because the measure
\eqn\measure{\int_0^\infty dp_+ \int_0^\infty dp_- \int d^{D-2} p_\perp 
\delta (p_+ p_- - p_\perp^2)} 
is Lorentz invariant. The left-hand side, however, contains
a derivative with respect to $x^0$, hence transforms in the same way as
$p_0$.

Now we perform a parallel computation in the
light-cone frame. Here, the number of particles radiated per unit
light-cone time is
\eqn\lcrate{ {dN\over d x^+} \sim 
\int_0^\infty dp_- \int_0^\infty dp_+ \int d^{D-2} p_\perp 
\delta (p_+ p_- - p_\perp^2)  
e^{-p_+/T_+} e^{-p_-/T_-} {\cal A} p_+
\ .}
The factor of $p_+$ is needed for correct boost invariance,
since the left-hand side contains a derivative with respect to $x^+$.
In the rest frame of the black hole, we have
\eqn\lctemp{T_+^{\rm rest}= T_-^{\rm rest}\sim T_H \sim 1/R_S \ .}
If we carry out a boost
\eqn\boost{ x^- = {R\over R_S} x^-_{\rm rest}\ ,
\qquad\qquad x^+ = {R_S\over R} x^+_{\rm rest}
\ ,}
then the new temperatures are
\eqn\boostemp{ T_-= T_-^{\rm rest} {R_S\over R} \sim 1/R\ , \qquad T_+=
T_+^{\rm rest} {R\over R_S}  \sim R/R_S^2
\ .}
Doing the integrals, we find
\eqn\semi{ {dN\over d x^+} \sim {\cal A} (T_+ T_-)^{(D-2)/2} T_+ \sim T_+
\sim R (G_D S)^{- 2/(D-2)}
\ .}

If we compactify $x^-$ on a circle of radius $R$, 
and work with $S\sim N$,
then the integral over $p_-$ is replaced by sum,
\eqn\discrete{ {dN\over d x^+} \sim {1\over R}\sum_{n>0}
\int_0^\infty dp_+  \int d^{D-2} p_\perp 
\delta \left ({n\over R} - {p_\perp^2\over p_+}\right )  
e^{-p_+/T_+} e^{-n/(R T_-)} G_D N
\ . 
}
While this changes the normalization, the result \semi\ still
holds.  Thus, our formula \rate\
for the rate of emission of D0-branes from
one of our matrix theory black holes coincides (up to uncalculated
numerical factors of order one) with the 
semiclassical Hawking evaporation rate in the
light cone frame, \semi.

\newsec{The Rate of Mass Loss}

Using arguments analogous to those that led to \proba, 
we may obtain from the matrix model a formula for
the rate of light cone energy loss per unit light cone time,
\eqn\enloss{{dE\over d x^+} \sim N {1\over R}\sum_{n>0}
\int_0^\infty dp_+  \int d^{D-2} p_\perp 
\delta \left ({n\over R} - {p_\perp^2\over p_+}\right )  
p_+ |A(n, y)|^2 
\ .}
Estimating the integral using the previously stated assumptions,
we find
\eqn\enlossb{{dE\over d x^+} \sim T_+^2 \sim {R^2\over R_S^4}
\ .}
Now using
\eqn\invariantmass{M^2 = 2 E {N\over R}\ ,
}
we have
\eqn\massloss{{d M\over dx^+} = {N\over M R} {dE\over d x^+} + 
{M\over 2 N} {dN\over d x^+}\sim {R\over R_S^3}\ .}
Boosting back to the rest frame, we obtain
\eqn\masslossb{
{dM\over d x^+_{\rm rest}}= {d M\over dx^+} {dx^+\over dx^+_{\rm rest}} 
\sim 1/R_S^2 \ ,}
where we used the fact that
\eqn\timdil{
{dx^+\over dx^+_{\rm rest}} = {R_S\over R}
\ .}
The Matrix theory result for the rate of mass loss in the rest
frame, \masslossb, is consistent with the standard semiclassical result.

What we have shown in this note is that plausible assumptions about
the D0-brane emission process from the metastable bound state describing
the black hole lead to the radiation rate consistent with the
semiclassical calculations, up to a constant of proportionality of order
1. Just as in the semiclassical analysis, suppression of the rate
for large entropy $N$ comes from the smallness of the one-particle
phase space available at an energy comparable to the Hawking temperature
($T_H$ scales as $N^{-1/(D-2)}$).
Our analysis should be regarded as a plausibility argument.
In particular, we need a microscopic argument for
why at low momenta
the square of the matrix element for liberating a D0-brane is
of order $G_D$. Nevertheless, we believe that we have described
the correct mechanism for evaporation of Schwarzschild
black holes in Matrix theory. 

\bigskip
\bigskip
\centerline{\bf ACKNOWLEDGEMENTS}
\bigskip

We are grateful to Lenny Susskind for important discussions and
to the Physics Department of Stanford University for its 
hospitality.
The work of T.B. was supported in part by the
Department of Energy under Grant No. DE - FG02
- 96ER40959 and that  of W.F. was supported in part by the Robert A.
Welch Foundation and by NSF Grant PHY-9511632.
The work of I.R.K was supported in part by the DOE grant
DE-FG02-91ER40671,
the NSF Presidential Young Investigator Award PHY-9157482, and the
James S.{} McDonnell Foundation grant No.{} 91-48.

\listrefs

\end